# A Generalised Deep Meta-Learning Model for Automated Quality Control of Cardiovascular Magnetic Resonance Images

Shahabedin Nabavi[1], Hossein Simchi[1], Mohsen Ebrahimi Moghaddam[1], Ahmad Ali Abin[1], Alejandro F. Frangi[2,3,4,5]

1- Faculty of Computer Science and Engineering, Shahid Beheshti University, Tehran, Iran.
2- Centre for Computational Imaging and Simulation Technologies in Biomedicine (CISTIB), The University of Leeds, Leeds LS2 9JT, U.K.
3- Leeds Institute for Cardiovascular and Metabolic Medicine (LICAMM), School of Medicine, University of Leeds, Leeds LS2 9JT, U.K.
4- Medical Imaging Research Center (MIRC), Cardiovascular Science and Electronic Engineering Departments, KU Leuven, Leuven, Belgium.
5- Alan Turing Institute, London, U.K.

**Corresponding Author:** Mohsen Ebrahimi Moghaddam (PhD)

**Address:** Faculty of Computer Science and Engineering, Shahid Beheshti University, Tehran, Iran.

**Email:** m_moghadam@sbu.ac.ir

**Phone:** +98 912 140 5308

**Conflicts of Interests:** There are no conflicts of interests to declare.

**Abstract**

**Background and Objectives:** Cardiovascular magnetic resonance (CMR) imaging is a powerful modality in functional and anatomical assessment for various cardiovascular diseases. Sufficient image quality is essential to achieve proper diagnosis and treatment. A large number of medical images, the variety of imaging artefacts, and the workload of imaging centres are among the things that reveal the necessity of automatic image quality assessment (IQA). However, automated IQA requires access to bulk annotated datasets for training deep learning (DL) models. Labelling medical images is a tedious, costly and time-consuming process, which creates a fundamental challenge in proposing DL-based methods for medical applications. This study aims to present a new method for CMR IQA when there is limited access to annotated datasets.

**Methods:** The proposed generalised deep meta-learning model can evaluate the quality by learning tasks in the prior stage and then fine-tuning the resulting model on a small labelled dataset of the desired tasks. This model was evaluated on the data of over 6,000 subjects from the UK Biobank for five defined tasks, including detecting respiratory motion, cardiac motion, Aliasing and Gibbs ringing artefacts and images without artefacts.

**Results:** The results of extensive experiments show the superiority of the proposed model. Besides, comparing the model's accuracy with the domain adaptation model indicates a significant difference by using only 64 annotated images related to the desired tasks.

**Conclusion:** The proposed model can identify unknown artefacts in images with acceptable accuracy, which makes it suitable for medical applications and quality assessment of large cohorts.

## 1- Introduction

Better clinical decision-making depends largely on medical imaging. Cardiovascular magnetic resonance (CMR) imaging has become a powerful diagnostic tool as a non-invasive, high spatial resolution modality. It is the gold standard in the evaluation of ventricular function. This modality has several clinical applications in analysing cardiac function and diagnosing heart disease [1]. The development of algorithms for CMR image analysis requires the extraction of robust features. Robust imaging features can improve prognostic, diagnostic and predictive accuracy. In this way, the availability of large-high-quality datasets is essential for extracting semantic and agnostic features from medical images [2]. Although population-based cohort studies such as the UK Biobank [3] have provided large-scale imaging datasets, image quality assessment (IQA) remains a challenge.

Image quality is directly related to the visual perception of the image. Respiratory and cardiac motions, fast blood flow, signal acquisition problems, etc., can lead to artefacts in CMR images and degrade image quality. These artefacts are artificial and obscure things in an image considered obstacles to a proper diagnosis. Many of these artefacts may appear in CMR images, including motion, Gibbs ringing, aliasing or wraparound, chemical-shift artefacts and those related to B0-inhomogeneities [4]. Assessing image quality subjectively in the large-scale studies like UK Biobank is a time-consuming, laborious, and non-reproducible task [5]. So, it is necessary to develop automated methods for evaluating image quality for these studies.

Training automated IQA models requires a large annotated dataset that specifies the artefacts that appeared in the images. In medical applications, access to such a dataset looks impossible. Medical images are not self-explanatory and need to be interpreted by experts. In such situations, the annotation of medical images due to the difference in the level of knowledge and the limitations of the human visual system needs to be done by many experts so the image labels can be trusted [6]. Performing such an operation in large datasets would be almost impossible and costly. It seems better to use machine learning methods that can handle the problem of not having access to large annotated datasets.

Machine learning models succeed in applications with bulk labelled data, while lack of access to such data prevents models from achieving high performance. Some machine learning approaches have been proposed to overcome this problem that can be more successful with less annotated data. Among these, methods such as few-shot learning [7], transfer learning [8], domain adaptation [9], and meta-learning [10] can be mentioned. These approaches can improve the performance of machine learning algorithms despite the lack of access to bulk labelled data. In the previous study [11], we addressed automated IQA in CMR images using an unsupervised domain adaptation method and introduced it as a baseline for not accessing bulk annotated data. In the current study, we propose a generalised optimisation-based deep meta-learning model for IQA in CMR images that can be superior to our previous research and more reliable for clinical applications.

The contributions of this study are:

(1) An automatic IQA model for CMR images with different views, including short-axis cine CMR and two, three and four-chamber long-axis CMR, is presented.
(2) A model trained to detect four common types of CMR artefacts: respiratory motion, cardiac movement, aliasing, and Gibbs. This model is able to address the lack of large annotated datasets.
(3) This study proposes an optimisation-based deep meta-learning model that is more generalised than earlier models. This proposed model is validated based on several experiments and can be used in future studies.

The rest of this article is organised as follows. Section 2 reviews previous studies on Deep Meta-Learning for IQA and Automated CMR IQA. Section 3 describes preparing data for different tasks, problem formulation and the proposed model. Section 4 is dedicated to the datasets descriptions, implementation details and how to evaluate the model by defining various experiments. Sections 5 and 6 present the results of the experiments and the discussion about the results, respectively. Finally, Section 7 deals with the conclusion of the study.

## 2- Related Works

### 2-1- Deep Meta-Learning for IQA

Meta-learning refers to the training of new tasks based on prior knowledge. Based on the experiences and knowledge gained from learning other tasks, called meta-knowledge, it is tried to make learning the new task faster. Increasing the speed of learning and improving the structure of machine learning approaches for the absence of data labels are among the benefits of meta-learning. In this learning method, previous experiences gained in similar tasks are used to learn new tasks in a knowledge-transfer manner [12]. Meta-learning approaches can be classified into three groups: model-, metric- and optimisation-based. In model-based techniques, recurrent neural networks are used to memorise knowledge gained from tasks to learn new tasks faster and more effectively. In metric-based methods, an attempt is made to learn an embedding function using meta-knowledge that can map each input to a suitable feature space to perform operations such as classification more efficiently. Optimisation-based methods use a bi-level optimisation process in which the parameters of a model are trained on a set of specific tasks, and fine-tuning the model on new tasks is then used for quick training with few training samples [10].

In IQA using meta-learning, several studies have used optimisation-based methods [13-15], and one research has leveraged meta-reinforcement learning [16]. In their first study, Zhu et al. [14] developed a deep regression network based on bi-level optimisation for IQA. Then, they added a spatial pyramid pooling module to their structure in their subsequent study [15]. Both of these studies show the superiority of the meta-learning method over earlier IQA methods. In Yan et al. [13] study, using a meta-learning method based on the stochastic gradient descent method has been considered, which also shows the superiority of meta-learning over some of the known IQA methods. The only study conducted in the medical IQA field is the study of [16], which used meta-reinforcement learning to evaluate the quality of trans-rectal ultrasound and chest X-ray images.

### 2-2- Automated CMR IQA

Assessing the quality of medical images is a challenging issue for various reasons. Due to the impossibility of accessing the quality reference image without distortion in medical IQA, many full-reference methods cannot be used for medical IQA. Therefore, it is necessary that the approaches in this field be no-reference. Besides, the artefacts in medical images are dramatically different due to the variety of imaging modalities and the anatomical diversity of the areas being imaged. The bulk of the data and the lack of access to artefact labels in images are other issues that challenge the medical image quality assessment. Thus, we need modality-specific methods that can automatically evaluate the quality of images with no reference [17].

Several studies have evaluated the quality of CMR images and focused exclusively on this modality's artefacts. Some of these studies [18-23] have considered complete heart coverage, particularly complete left ventricular coverage, as primary quality control. In these studies, techniques such as decision forest [19, 20] and DL models [21-23] have investigated complete heart coverage in the UK Biobank. Motion artefacts [19, 20, 24] and contrast estimation [19, 20] are other items that the proposed models have evaluated. However, due to the importance of CMR IQA, models should cover a broader range of common artefacts. In our previous study [11], we addressed four types of common artefacts in these images. In that study, we also tried to address the domain adaptation approach to overcome the shortage of annotated data. Despite the studies done, this field still has great potential for further research.

## 3- Method

This section deals with preparing the required data and tasks and describes the proposed deep meta-learning model for CMR IQA when there is no convenient access to large annotated datasets. First, the problem is formulated, and then the structure of a meta-learning baseline model is explained. Finally, an improved model with more generality than the baseline model is proposed.

### 3-1- Data Preprocessing

Short- and long-axis CMR images from over 6,000 subjects of the UK Biobank and 33 subjects of the CMR imaging database of York University (YU) [25] are used to prepare the datasets for this study. Since to meet the goals of this study, there is a need for bulk annotated data, and this data is not available, so the technique of adding synthetically but realistic artefacts using k-space manipulation is used. Previous studies [24, 26] have added artefacts to CMR images by k-space manipulation techniques to generate datasets for CMR IQA. Therefore, considering that respiratory and cardiac motion, Gibbs, and aliasing artefacts are the most common artefacts in CMR images [4, 27], we used the techniques introduced in our previous study [11] to add these artefacts to images and generate synthetically but realistic artefacted images.

Reference images from the mentioned datasets are first subjected to a 1-dimensional translation to generate degraded images by respiratory motion artefact. The reference and translated images are then transferred to the k-space using the Fourier transform. The k-space lines of these two images are combined based on a sinusoidal pattern to form the degraded k-space, which is reconstructed to generate the degraded image with the respiratory motion artefact. For degraded images with the cardiac motion artefact, the k-space lines of the short-axis images must be combined in a temporal sequence. Also, the ideal low-pass filtering of the k-space corresponding to the reference image can create Gibbs distortion in CMR images. Besides, images degraded by aliasing can be achieved by alternately eliminating k-space lines of reference images. Details on how to add these artefacts to the images and related references are described in our previous study [11].

By adding these four mentioned artefacts with the k-space manipulation techniques, five classes are considered for the experiments required in this study, including images with respiratory motion, cardiac motion, Gibbs and Aliasing artefacts and artefact-free images. 40,960 images are used for each class to balance these five classes. Therefore, five categories are formed to define five tasks in this study.

### 3-2- Problem Formulation

Suppose $\mathcal{T} = \{\mathcal{T}_c\}$ is a set of tasks where $c \in \{Respiratoy\ Motion, Cardiac\ Motion, Gibbs, Aliasing, Artefact\_Free\}$ indicates a task involving one type of artefact identification or no artefact detection in CMR images. Given the purpose of the study, suppose that $S_{Artefact\_Specific}$ contains samples in which the type of artefact is known, meaning they are labelled. This set is to be used to learn the prior model to generate meta-knowledge. $S_{Artefact\_Specific}$ is represented by Equation 1.

$$\mathcal{S}_{Artefact\_Specific} = \{(x_i, y_i)\}_{i=1}^{N}, y_i \in \{1, 2, ..., l\} \quad (1)$$

where $x_i$ represents an image, $y_i$ shows the image label indicating the artefact type, and $l$ is the number of classes used in the prior model training phase.

Suppose $\mathcal{S}_{Unseen\_Artefact}$ is another set of data used as the target set. The target set consists of tasks used to identify artefacts in CMR images that are unknown. The set $\mathcal{S}_{Unseen\_Artefact}$ is expressed as Equation 2.

$$\mathcal{S}_{Unseen\_Artefact} = \{\mathcal{S}_{Unseen\_Artefact}^{Labelled}, \mathcal{S}_{Unseen\_Artefact}^{Unlabelled}\} \quad (2)$$

In this dataset, there are images whose artefact type is unseen ($\mathcal{S}_{Unseen\_Artefact}^{Unlabelled}$), and only a tiny portion of this dataset has an artefact-type label ($\mathcal{S}_{Unseen\_Artefact}^{Labelled}$). $\mathcal{S}_{Unseen\_Artefact}^{Labelled}$ is used to fine-tune the prior model to identify unseen artefacts. $\mathcal{S}_{Unseen\_Artefact}^{Labelled}$ and $\mathcal{S}_{Unseen\_Artefact}^{Unlabelled}$ are shown in Equations 3 and 4.

$$\mathcal{S}_{Unseen\_Artefact}^{Labelled} = \{(x_i', y_i')\}_{i=1}^{N}, y_i' \neq y_i \quad (3)$$

$$\mathcal{S}_{Unseen\_Artefact}^{Unlabelled} = \{x_i''\}_{i=1}^{N} \quad (4)$$

where $x_i'$ is an image with a $y_i'$ label, the $y_i'$ is different from the $y_i$. Also, $x_i''$ does not have an artefact type label.

### 3-3- A Generalised Deep Meta-Learning Model for CMR IQA

The overview of the proposed model of this study is shown in Figure 1. Our approach proceeds as follows:

- Train and test datasets preparation
- Prior model meta-training on artefact-specific tasks
- Prior model fine-tuning for unseen artefacts

First, it is necessary to prepare the data for training and testing the proposed model. The actions taken at this stage of the proposed model are important because they play a significant role in further generalisation. The model is then trained on the prepared data to obtain the parameters of the prior model or the meta-knowledge. In the final step, the model is fine-tuned on $\mathcal{S}_{Unseen\_Artefact}^{Labelled}$. The final parameters of the model are used for testing on $\mathcal{S}_{Unseen\_Artefact}^{Unlabelled}$. The proposed deep meta-learning model for CMR IQA is described in Algorithm 1.

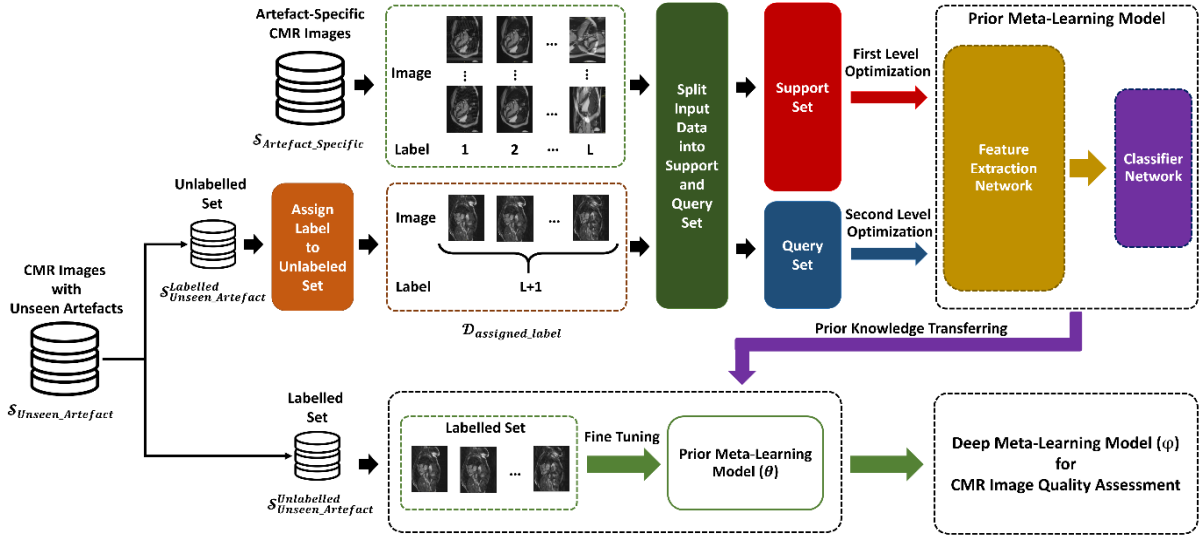

Figure 1: Overview of the proposed deep meta-learning model for CMR IQA.

### 3-3-1- Train and Test Datasets Preparation

In the existing studies, tasks are used to prepare the training data needed to learn the prior model that the task is known. Thus, we use $\mathcal{S}_{Artefact\_Specific}$ as the training data for the prior model. In comparison, the difference between the proposed method and the existing meta-learning models is that in addition to $\mathcal{S}_{Artefact\_Specific}$, it also uses $\mathcal{S}^{Unlabelled}_{Unseen\_Artefact}$ to generate the training dataset. For this purpose, assuming there are $l$ tasks in $\mathcal{S}_{Artefact\_Specific}$, we assign a temporary label called $l+1$ to all data in set $\mathcal{S}^{Unlabelled}_{Unseen\_Artefact}$ and make $\mathcal{D}_{assigned\_label}$ as follows.

$$\mathcal{D}_{assigned\_label} = \{(x_i'', l+1)\}_{i=1}^N \quad (5)$$

Now to create the meta-training set for learning the prior model, we make set $\mathcal{D}_{meta\_train}$ by Equation 6.

$$\mathcal{D}_{meta\_train} = \mathcal{S}_{Artefact\_Specific} \cup \mathcal{D}_{assigned\_label} \\ = \{(x_i^{meta}, y_i^{meta})\}_{i=1}^N, y_i^{meta} \in \{1, 2, \ldots, l, l+1\} \quad (6)$$

To perform bi-level optimisation in prior model training, $\mathcal{D}_{meta\_train}$ must be divided into $\mathcal{D}_{meta\_train}^{support}$ and $\mathcal{D}_{meta\_train}^{query}$. After training the prior model on $\mathcal{D}_{meta\_train}$ and generating meta-knowledge, the resulting model is fine-tuned on $\mathcal{S}^{Labelled}_{Unseen\_Artefact}$ to prepare the final model for testing. Finally, the model is evaluated on $\mathcal{S}^{Unlabelled}_{Unseen\_Artefact}$, which includes tasks whose artefact types are unknown. Therefore, the dataset used to test the model is $\mathcal{D}_{test}$, which is shown in Equation 7.

$$\mathcal{D}_{test} = \mathcal{S}^{Unlabelled}_{Unseen\_Artefact} \quad (7)$$

### 3-3-2- Prior Model Meta-Training on Artefact-Specific Tasks

The main idea of optimisation-based meta-learning methods is to learn the model parameters so they can be quickly adapted to similar new tasks. Considering this issue, we want to gain prior or meta-knowledge of known artefacts so this knowledge has the proper generalisation to learn the task of unknown artefact detection rapidly. To generate meta-knowledge, we have a classification problem for artefact-specific tasks. This problem is defined:

$$f_\theta: x_i^{meta} \to \hat{y}_i^{meta} \tag{8}$$

where $\theta$ is the initial parameters of the model, $x_i^{meta}$ is the $i^{th}$ input image, and $\hat{y}_i^{meta}$ is the $i^{th}$ output obtained from the classification process. The initial parameters $\theta$ are considered as the inductive bias of the model, which means the set of assumptions that the learning model uses to predict samples that it has not seen.

The goal in this stage is to train the model to achieve the best performance in the least number of optimisation steps for each task defined in the meta-training stage. For this purpose, one should seek to reduce the loss function of the prior model, which is expressed:

$$\mathcal{L}_j(\hat{y}^{meta}, y_j^{meta}) = -\frac{1}{N_j}\sum_{i=1}^{N_j} y_{ij}^{meta} \cdot \log(\hat{y}_i^{meta}) \tag{9}$$

where $j \in \mathcal{T}$ shows the task used in training the prior model, $N_j$ and $i$ represent the number of data and the $i^{th}$ instance of a particular task, and $y^{meta}$ indicates the corresponding ground truth.

By selecting a mini-batch from the data in $\mathcal{D}_{meta\_train}^{support}$ and calculating the error of the feedforward process for a particular task using Equation 9, the model parameters are updated using the Adam optimiser [28] in Equation 10.

$$\theta'_j \leftarrow Adam\_Optimizer\left(\mathcal{L}_j^{support}(\hat{y}^{meta}, y^{meta}), \theta\right) \tag{10}$$

We expect the $\theta'_j$ model obtained from training on the support set for a particular task $j$ to perform well on the query set. Therefore, $\theta'_j$ model parameters are updated on $\mathcal{D}_{meta\_train}^{query}$ in Equation 11.

$$\theta_j \leftarrow Adam\_Optimizer\left(\mathcal{L}_j^{query}(\hat{y}^{meta}, y^{meta}), \theta'_j\right) \tag{11}$$

Equation 12 should obtain the final parameters of the model for all the tasks in the prior stage.

$$\theta \leftarrow \theta - \eta \left(\frac{1}{k}\sum_{j=1}^{k}(\theta - \theta_j)\right) \tag{12}$$

where $\eta$ is used as the learning rate for the outer update of the prior model, and $k$ shows the number of tasks in the prior stage.

By repeating the above steps for different mini-batches and finishing the training process of the prior model using this bi-level optimisation method, we obtain meta-knowledge on the tasks of artefact detection in CMR images. This meta-knowledge is the input of the third step, fine-tuning the prior model. The prior model consists of two general parts: feature extraction and classifier networks. A ResNet18 [29] network pre-trained on ImageNet is used for feature extraction in the prior model. The classifier network consists of three modules; each contains a fully connected layer, batch normalisation and ReLU function. The modules have 1024, 512 and as many neurons as the tasks of the prior stage in their fully connected layer, respectively.

### 3-3-3- Prior Model Fine-tuning for Unknown Artefacts

At this stage, the meta-knowledge obtained from prior model training should be fine-tuned for the data in which the artefact type is unknown. $S^{Labelled}_{Unseen\_Artefact} = \{(x'_i, y'_i)\}^N_{i=1}, y'_i \neq y_i$ is used for fine-tuning. For the fine-tuning of the prior model, Equation 13 is the loss function.

$$\mathcal{L}_{fine\_tune}(\hat{y}', y') = -\frac{1}{N}\sum_{i=1}^{N}\sum_{j=1}^{M} y'_{ij} \cdot \log(\hat{y}'_{ij}) \tag{13}$$

where $N$ represents the number of samples in $S^{Labelled}_{Unseen\_Artefact}$ and $M$ indicates the number of unknown tasks. Then the Adam optimiser is used to update the prior model and create the fine-tuned model. Equation 14 shows this update process.

$$\varphi \leftarrow Adam\_Optimizer(\mathcal{L}_{fine\_tune}(\hat{y}', y'), \theta) \tag{14}$$

$\varphi$ will be the final model leveraged for the quality assessment of CMR images with unknown artefacts. Finally, dataset $S^{Unlabelled}_{Unseen\_Artefact}$ is used to evaluate the $\varphi$ model to determine the efficiency of the final model.

---

**Algorithm 1** A Generalised Deep Meta-Learning Model for CMR IQA

**Require:**
- ▷ $S_{Artefact\_Specific} = \{(x_i, y_i)\}^N_{i=1}, y_i \in \{1, 2, ..., l\}$, where $x_i$ is an image, $y_i$ is the image label/artefact type, and $l$ is the number of classes used in the prior model training
- ▷ $S^{Labelled}_{Unseen\_Artefact} = \{(x'_i, y'_i)\}^N_{i=1}, y'_i \neq y_i$, where $x'_i$ is an image with a $y'_i$ label
- ▷ $S^{Unlabelled}_{Unseen\_Artefact} = \{x''_i\}^N_{i=1}$, where $x''_i$ is an image without label
- ▷ $\theta$ : Initial parameters of prior model pre-trained on ImageNet
- ▷ $\eta$ : Outer learning rate

**Ensure:**
- ▷ $\varphi$ : Parameters of Deep Meta-Learning Model for CMR IQA

1: $\mathcal{D}_{assigned\_label} \leftarrow \{(x''_i, l+1)\}^N_{i=1}$     ▷ Assign Label to Unlabelled Set
2: $\mathcal{D}_{meta\_train} \leftarrow S_{Artefact\_Specific} \cup \mathcal{D}_{assigned\_label} = \{(x^{meta}_i, y^{meta}_i)\}^N_{i=1}$, $y^{meta}_i \in \{1, 2, ..., l, l+1\}$

    ▷ Prior Model Meta–Training on Artefact–Specific Tasks :
3: **for** $iteration = 1, 2, ...$ **do**
4:    Select a mini-batch from $\mathcal{D}_{meta\_train}$
5:    **for** All $x^{meta}_i \in$ mini-batch **do**
6:      $\hat{y}^{meta}_i \leftarrow f_\theta(x^{meta}_i)$
7:    **end for**
8:    **for** $j = 1, 2, ..., l$ **do**
9:      $\theta'_j \leftarrow$ Adam_Optimizer $(\mathcal{L}^{support}_j(\hat{y}^{meta}, y^{meta}_j), \theta)$
10:     $\theta_j \leftarrow$ Adam_Optimizer $(\mathcal{L}^{query}_j(\hat{y}^{meta}, y^{meta}_j), \theta'_j)$
11:    **end for**
12:    $\theta \leftarrow \theta - \eta(\frac{1}{l}\sum_{j=1}^{l}(\theta - \theta_j))$
13: **end for**

    ▷ Prior Model Fine–tuning for Unknown Artefacts :
14: **for** $x'_i \in S^{Labelled}_{Unseen\_Artefact}$ **do**
15:    $\hat{y}'_i \leftarrow f_\theta(x'_i)$
16: **end for**
17: $\varphi \leftarrow$ Adam_Optimizer$(\mathcal{L}_{fine\_tune}(\hat{y}', y'), \theta)$

18: **return** $\varphi$     ▷ Parameters of Deep Meta–Learning Model for CMR IQA

## 4- Experimental Setup

### 4-1- Dataset Description

This study uses two datasets of CMR images, including the UK Biobank and the YU dataset. Short- and long-axis CMR images of over 6,000 subjects of the UK Biobank are used to create diversity in the input data for training and evaluation of the proposed model. Also, all the short- and long-axis CMR images of 33 participating subjects in the YU dataset are leveraged to compare the proposed model with an unsupervised domain adaptation technique.

UK Biobank's CMR imaging infrastructure is a 1.5 Tesla scanner with 48 receiver channels and a 45 $mT/m$ and 200 $T/m/s$ gradient system. For imaging, an 18-channel anterior body surface coil is combined with 12 elements of an integrated 32-element spine coil. Electrocardiogram (ECG) gating is leveraged for cardiac synchronisation. 2D cine balanced steady-state free precession (bSSFP) short-axis and long-axis images have in-plane spatial resolution 1.8 × 1.8 $mm$, slice thickness 8 $mm$ and slice gap 2 $mm$. To obtain more details about this dataset, refer to [30].

7,980 short- and long-axis images have been acquired using a GE Genesis Signa MR scanner. Fast imaging employing steady-state acquisition (FIESTA) scanning protocol has been used. In the short-axis view, the number of frames is 20; in the long-axis view, the number of slices varies between 8 and 15. The spatial resolution of slices is 256 × 256, and the slice gap is between 6 and 13 $mm$. In this study, under 18-year-old subjects have been examined. For more information, it is recommended to read [25].

### 4-2- Model Training and Implementation Details

According to its structure, the training of the proposed model is done in two stages: prior and fine-tuning. The loss function defined in Equation 9 and Adam optimiser are used to train the prior model. The training samples for each class and the learning rate in the prior stage are 10,000 and 0.0001, respectively. For training in the fine-tuning stage, the loss function of Equation 13 and Adam optimizer are used. The total of training and testing data at this stage is 40,960 images for each class, and the learning rate is equal to 0.0001. The training process for both stages is done using cross-validation with a batch size of 64 in 30 epochs. All the codes related to this study were developed using the PyTorch package [31] in the Python environment. The choice of hyper-parameters is based on ablation studies. The computer system used to train and evaluate the proposed models has an NVIDIA GeForce GTX TITAN X GPU and 16GB of RAM.

### 4-3- Experimental Design and Evaluation Metrics

Several experiments were designed and started to evaluate the proposed model. In these tests, the size of $S_{Unseen\_Artefact}^{Labelled}$ and the type of artefacts examined are changed in the fine-tuning stage to assess the proposed model's generality. In these experiments, three implementation modes are considered to make a better comparison:

- In the first mode, the baseline case is evaluated, so the prior model is trained on the $S_{Artefact\_Specific}$ and $S_{Unseen\_Artefact}^{Labelled}$ and tested on $S_{Unseen\_Artefact}^{Unlabelled}$. In this case, the concept of meta-learning is not leveraged.
- In the second mode, the proposed model is utilised without using the phase of assigning label to $S_{Unseen\_Artefact}^{Unlabelled}$ data. In this way, only $S_{Artefact\_Specific}$ is used in prior model training, line 1 of algorithm 1 is eliminated, and $\mathcal{D}_{assigned\_label}$ is assumed to be null.
- In the third mode, all the steps mentioned in the proposed algorithm are executed.

The size of the set $S_{Unseen\_Artefact}^{Labelled}$ in the fine-tuning stage is 64, 128 and 256 for each artefact type in three runs. The artefacts evaluated for the fine-tuning stage have been changed in four experiments. Besides, to compare with previous studies, the only developed model to overcome the lack of access to annotated data is compared under fair conditions with the proposed model to determine the improvement rate.

Four metrics, including accuracy (ACC), precision (PR), recall (RE), and F-measure [32], are run to evaluate the proposed model.

## 5- Results

### 5-1- Analysis of the Proposed Model

The analysis process of the proposed model was done in four experiments, considering $S_{Artefact\_Specific}$, $S_{Unseen\_Artefact}^{Labelled}$, and $S_{Unseen\_Artefact}^{Unlabelled}$ sets with different combinations of five types of studied artefacts. Each experiment is repeated three times with varying sizes of the $S_{Unseen\_Artefact}^{Labelled}$ set consisting of 64, 128 and 256 images of each class of the fine-tuning stage. Therefore, we have four experiments, each repeated three times with different sizes of $S_{Unseen\_Artefact}^{Labelled}$ to enable comparisons and determine the effectiveness of the number of annotated data in the fine-tuning stage. It should be mentioned that we have three implementation modes, which were explained in Section 4-3. Therefore, 12 executions related to the above four experiments are repeated for each mode to provide the possibility of comparing the proposed model's generality. The experimental results and the combination of artefacts related to each test are shown in Table 1. When the number of samples of the classes is precisely balanced, the weighted recall value is equal to the accuracy [33]. Thus, the recall values were not entered in the table below.

Table 1: The results of experiments based on accuracy, precision and F-measure metrics. The numbers in parentheses are standard deviations. 64, 128 and 256 indicate the number of annotated data in each class in $S_{Unseen\_Artefact}^{Labelled}$. M1, M2 and M3 show the three defined implementation modes.

|  |  |  | Accuracy (%) | | | Precision (%) | | | F-measure (%) | | |
|---|---|---|---|---|---|---|---|---|---|---|---|
|  |  |  | M1 | M2 | M3 | M1 | M2 | M3 | M1 | M2 | M3 |
| Experiment 1 | Prior Classes:<br>- Respiratory Motion<br>- Gibbs<br>- Artefact free<br><br>Fine-tuning Classes:<br>- Aliasing<br>- Cardiac Motion | 64 | 51.18 (3.46) | 78.52 (2.97) | 78.95 (15.71) | 34.33 (17.69) | 80.80 (2.98) | 84.46 (10.84) | 39.34 (11.37) | 79.64 (2.87) | 81.42 (13.41) |
|  |  | 128 | 50.81 (3.44) | 79.24 (1.69) | 87.69 (13.42) | 33.18 (17.12) | 81.69 (2.05) | 91.42 (8.32) | 38.43 (10.70) | 80.44 (1.71) | 89.40 (11.08) |
|  |  | 256 | 50.50 (2.00) | 79.53 (2.19) | 99.28 (3.57) | 32.16 (16.23) | 81.71 (2.54) | 99.46 (2.37) | 37.66 (9.72) | 80.60 (2.25) | 99.36 (3.01) |
| Experiment 2 | Prior Classes:<br>- Respiratory Motion<br>- Cardiac Motion<br>- Artefact free<br><br>Fine-tuning Classes:<br>- Aliasing<br>- Gibbs | 64 | 50.70 (2.54) | 60.07 (3.92) | 88.50 (11.30) | 34.75 (17.88) | 66.76 (5.61) | 91.15 (7.64) | 39.30 (10.85) | 63.12 (3.87) | 89.73 (9.58) |
|  |  | 128 | 50.20 (1.08) | 59.87 (4.20) | 85.61 (16.91) | 31.55 (15.93) | 66.19 (6.02) | 89.80 (11.00) | 37.10 (9.04) | 62.76 (4.27) | 87.45 (14.24) |
|  |  | 256 | 50.64 (2.29) | 59.82 (4.35) | 90.14 (12.67) | 33.66 (17.12) | 65.91 (6.23) | 92.68 (8.10) | 38.65 (10.39) | 62.58 (4.28) | 91.28 (10.58) |
| Experiment 3 | Prior Classes:<br>- Respiratory Motion<br>- Aliasing<br>- Artefact free<br><br>Fine-tuning Classes:<br>- Cardiac Motion<br>- Gibbs | 64 | 50.33 (1.59) | 71.11 (5.09) | 86.50 (14.48) | 35.01 (18.69) | 77.44 (2.50) | 90.56 (9.51) | 39.11 (10.65) | 74.08 (3.65) | 88.28 (12.25) |
|  |  | 128 | 50.58 (2.54) | 71.44 (5.82) | 96.76 (8.74) | 34.90 (17.68) | 77.40 (2.33) | 97.69 (5.14) | 39.38 (10.75) | 74.23 (4.13) | 97.17 (7.18) |
|  |  | 256 | 50.31 (1.53) | 72.00 (5.72) | 96.82 (7.53) | 35.34 (18.13) | 77.08 (3.06) | 97.58 (4.88) | 39.42 (10.53) | 74.41 (4.35) | 97.16 (6.31) |
| Experiment 4 | Prior Classes:<br>- Respiratory Motion<br>- Artefact free<br><br>Fine-tuning Classes:<br>- Aliasing<br>- Cardiac Motion<br>- Gibbs | 64 | 34.14 (1.97) | 50.07 (5.07) | 56.51 (9.52) | 17.87 (10.36) | 48.04 (12.05) | 55.43 (16.56) | 22.11 (7.80) | 48.45 (8.06) | 55.06 (11.79) |
|  |  | 128 | 33.98 (1.88) | 50.1 (4.29) | 61.31 (9.99) | 17.26 (10.02) | 48.27 (11.33) | 57.40 (16.39) | 21.60 (7.62) | 48.61 (7.21) | 58.54 (12.07) |
|  |  | 256 | 33.69 (1.15) | 50.74 (4.21) | 63.61 (10.06) | 16.48 (10.06) | 49.09 (11.25) | 63.92 (17.73) | 20.79 (7.22) | 49.24 (6.83) | 63.02 (13.20) |

### 5-2- Proposed Model versus Other Related Methods

An experiment was designed to compare the current study with the only previous study [11] that addressed the challenge of limited access to annotated data in CMR IQA. In that study, an unsupervised domain adaptation approach was used for the feasibility of training the model on a dataset and testing the model on a different dataset. Its purpose is to learn knowledge from a labelled source dataset and transfer it to use the model in another unlabelled target dataset. The datasets consisting of aliasing and Gibbs classes are used to perform this test so that the UK Biobank and YU are considered the source and target sets, respectively. The number of samples of each class in both datasets is 10,240. Training the model [11] is done using the labelled data from the UK Biobank dataset and 64 annotated images of the YU as a source set and the rest of the unlabelled samples of YU as the target set. Using 64 labelled samples from the YU dataset makes the execution semi-supervised, establishing fairness in the comparisons. Finally, the evaluation process is conducted on unlabelled YU samples.

The proposed model is also trained on aliasing and Gibbs classes from UK Biobank for prior model meta-training. Then, with the same 64 labelled samples of YU, the prior model fine-tuning process is performed. The proposed model evaluation in this experiment is conducted on the unlabelled part of the YU dataset. The results of the comparisons are presented in Table 2.

In addition to comparing with a CMR IQA method based on domain adaptation, the proposed method was compared with three well-known optimization-based meta-learning models. The results of these comparisons are also presented in Table 2.

Table 2: Comparison of proposed against an alternative model and three optimization-based meta-learning models. Standard deviations are in parentheses.

| Metric / Model | Accuracy (%) | Precision (%) | Recall (%) | F-measure (%) |
|---|---|---|---|---|
| **Unsupervised Domain Adaptation [11]** | 50.05 (0.67) | 25.33 (4.12) | 50.05 (0.67) | 33.54 (2.62) |
| **Model-Agnostic Meta-Learning (MAML) [34]** | 53.69 (1.53) | 54.09 (2.34) | 53.69 (1.53) | 53.89 (2.06) |
| **Antoniou et al. [35]** | 54.21 (3.72) | 55.75 (5.37) | 54.21 (3.72) | 54.97 (4.11) |
| **Reptile [36]** | 56.48 (4.42) | 58.33 (6.93) | 56.48 (4.42) | 57.39 (5.56) |
| **Proposed Model** | **61.14 (9.71)** | **65.43 (17.57)** | **61.14 (9.71)** | **62.23 (13.56)** |

## 6- Discussion

This study investigates the possibility of CMR IQA with different views where accessing annotated data is limited. Based on what was mentioned earlier, data labelling is time-consuming, costly, and laborious. On the other hand, bulk annotated data is required to train DL models so that overfitting does not occur. Therefore, we proposed a method that can achieve an acceptable level of generality with minimal labelled data. The proposed model was analysed with extensive tests, and based on the results, it can work promisingly in the lack of access to data labels. This model is also superior to previous models.

In the proposed method, we want to obtain prior knowledge using known distortions and generalise it on unknown distortions. This approach can make it possible to achieve promising accuracy in deep neural networks in the absence of access to bulk annotated data. By obtaining more generality, this method has increased the final accuracy so the results are much more reliable for medical applications than related studies [11]. Also, the proposed method can evaluate the quality of images of large datasets such as the UK Biobank, so there is no need for an extensive and time-consuming labelling process.

Extensive experiments were designed and executed to prove the proposed method's superiority. The analysis of these experiments shows that in the condition of not using any meta-learning method (mode 1), the model's training becomes random. This problem is seen in all experiments and shows that the model's performance is unacceptable, even with several labelled samples. However, the results are significantly improved in mode 2 compared to mode 1. Meanwhile, increasing the number of annotated samples for the baseline model fine-tuning (mode 2) makes little difference in the results and is mostly ineffective.

Compared to modes 1 and 2, the proposed method (mode 3) has the best results in all cases. Besides, increasing the number of annotated samples in mode 3 leads to improved outcomes, except in the second experiment. The random selection of annotated samples seems to be the reason for the pattern difference in the second experiment. It means that the random selection of labelled samples could not represent the proper distribution of the total data and various intensities of the artefacts. From experiments 1 and 3, it can be understood that according to the almost similar pattern of blurring and ghosting resulting from the respiratory and cardiac motion artefacts, if the detection of one of these two artefacts is the prior task and the other as the fine-tune task, it can increase the accuracy of the proposed model. By examining the results of experiment 2, this issue becomes more apparent. In this experiment, detecting both respiratory and cardiac motion artefacts is considered two prior tasks, and detecting two artefacts with different manifestations is regarded as two fine-tuning tasks. This difference in the artefacts' manifestations has caused a decrease in the accuracy in experiment 2 compared to experiments 1 and 3.

In experiment 4, when two tasks are considered for the prior stage and three tasks for the fine-tuning stage, again, the results of mode 1 show that not using the proposed method results in the random performance of the classifier. However, using meta-learning leads to the improvement of the results, and the results reach the highest level using the proposed method. Although the respiratory motion artefact is the only artefact detection task in the prior stage, the performance of the proposed method is noteworthy. Comparing the results with the domain adaptation method presented [11], MAML [34], Antoniou et al. [35], and Reptile [36] methods shows the superiority of the proposed optimisation-based meta-learning method. Besides, the simultaneous highness of all four metrics, including accuracy, precision, recall and F-measure, shows the remarkable performance of the proposed model in detecting unknown artefacts.

## 7- Conclusion

This study proposed a generalised optimisation-based deep meta-learning model that can evaluate the quality of CMR images even without access to the bulk annotated dataset. By training on the data related to artefact detection tasks in CMR images, from which there are enough labelled data, this model can provide the ability to identify other artefacts with a tiny amount of labelled data with promising accuracy. The proposed model was analysed and investigated through experiments on four common artefacts in CMR images, including respiratory and cardiac motion, aliasing and Gibbs ringing. The results show the superiority of this model over earlier models, and this model can be used for IQA in large cohorts. In future studies, we intend to examine more artefacts and make it possible to use k-space features in the CMR IQA.

**Acknowledgement**

This study has been conducted using the UK Biobank CMR dataset under Application 11350.